\documentclass[aps,prl,superscriptaddress,reprint]{revtex4-2}
\usepackage{graphicx}
\usepackage{xcolor}
\usepackage{hyperref}
\hypersetup{colorlinks=brown,linkcolor=RoyalBlue,citecolor=red,urlcolor=RoyalBlue}

\def\aap{Astron.\ Astrophys.\ }

	\def\apj{Astrophys.\ J.\ }

	\def\mnras{Mon.\ Not.\ R.\ Astron.\ Soc.\ }

	\def\prd{Phys.\ Rev.\ D }
	
	\def\prl{Phys.\ Rev.\ Lett.\ }

\begin{document}
\title{Accurate  and Unbiased Reconstruction of CMB B Mode using Deep Learning
}
\author{Srikanta Pal}
\email{srikanta18@iiserb.ac.in}
\affiliation{Department of Physics, Indian Institute of Science Education and Research, Bhopal, India}
\author{Sarvesh Kumar Yadav}
\email{sarvesh@rrimail.rri.res.in}
\affiliation{Raman Research Institute, C. V. Raman Avenue, Bengaluru, India}
\author{Rajib Saha}
\email{rajib@iiserb.ac.in}
\affiliation{Department of Physics, Indian Institute of Science Education and Research, Bhopal, India}
\author{Tarun Souradeep}
\email{tarun@rri.res.in}
\affiliation{Raman Research Institute, C. V. Raman Avenue, Bengaluru, India}

\begin{abstract}
An ingeniously designed autoencoder (PrimeNet) using simulated  observations of future generation ECHO satellite mission recovers CMB B mode map, angular spectrum for multipoles $\ell \lesssim 9$ and tensor to scalar ratio $r$ {\it limited only by cosmic variance down to $r= 0.0001$ and below}. We use diverse, realistically complex and detailed  foreground models. PrimeNet predicts accurate results even when  data with $r=0$ are  tested which were not used in training, implying robust and efficient predictive power. The work eliminates a major bottleneck of weak CMB B mode reconstruction and takes  a leap forward for understanding fundamental physics of the primordial Universe.

\end{abstract}

\maketitle


{\bf Introduction:} The general theory of relativity predicts that the tiny ripples that occur due to the disturbances in the energy and momentum flow lead to propagating gravitational wave and further influence the events taking place inside the space-time manifold~\cite{EINSTEIN1937}. A relic of these waves is predicted to be generated  ~\cite{Starobinski1979,RUBAKOV1982} in  a very early epoch of almost exponential expansion~\cite{Guth1981,Linde1982} of spatial scales, for a very brief interval of time.

This primordial gravitational wave (PGW) is one of the founding pillars of the cosmological inflation theory and serves as a unique probe to the physics of the  very early universe.
The direct estimation of PGW propagating since the inflation has not been possible till date as the signal could be extremely weak. They however leave their imprint on cosmic microwave background (CMB)~\cite{FABBRI1983,Abott1984, Lyth1984, Krauss1992}.  The PGW redistributes the space time metric and hence influences the propagation of the electromagnetic radiation. The primordial B mode component  of CMB  polarization over the large angular scales of the sky uniquely encodes the PGW or tensor modes of the metric components and therefore serves as the unambiguous  probe for detection of the signal~\cite{Seljak1997, Kamionkowski1997}. PGW  is quantified by the variable $r$, the so called ratio of tensor to scalar mode power spectrum.

Measuring PGW through the CMB B mode constrains physics of energy scales which are many orders of magnitude beyond the reach of any modern day terrestrial accelerators.   The tensor to scalar ratio constrains expansion rate of the universe at inflationary epoch when a scale of interest in density perturbations of the scalar field escapes the horizon and therefore directly constrains energy densities of such early times.  A fundamental problem in the inflationary models is the numerical value of the net change of the scalar field~\cite{Krauss2010}. If this value is larger than the Planck mass one can put major constraints on the theory of quantum gravity since one is then probing physics in an environment where manifestation of quantum gravitational effects becomes important.

Recent observations of Planck-BICEP~\cite{BICEP2022} constrain only an upper limit $r \lesssim 0.030$ at $k=0.005$ $\textrm{Mpc}^{-1}$. {\it In this work we focus on the fundamental problem of reconstruction of $r$.} We develop  a machine learning architecture (PrimeNet) for accurate and robust prediction of CMB B mode maps by training the network with simulations that contain realizations of CMB maps with discretely varying $r$ in between $0.0001$ and $0.055$. Each training realization represents  simulated full sky B mode maps of all $20$ frequency channels  of future generation satellite based CMB polarization mission `Exploring Cosmic Origin and History' (ECHO)~\cite{ECHO2022, ECHO2023} (to be placed  at the $\textrm{L}_2$ Lagrange point of Sun-Earth system by `Indian Space Research Organization' (ISRO)) with  realistically complex and diverse models of foregrounds and detector noise.   After the learning phase PrimeNet produces accurate estimates of the full sky CMB B mode signal during testing  in which $r$ was set as low as a value of zero suggesting a robust understanding of the network to efficiently remove the contamination even in those cases in which only lensing signal was present without any primordial B mode signal. Besides ECHO a number of other polarization projects~\cite{LiteBird2023, CMBS42022, SO2022} are also in active consideration.

Reconstruction of CMB temperature anisotropies from observations has been extensively studied in the literature~\cite{Tegmark2003, Delabrouille2003, Saha2006,Hinshaw2007, Eriksen2007, Saha2008, Eriksen2008, GNILC2011, Basak2013, Saha2011, Pietrobon2012, Delabrouille2012, Saha2016, Sudevan2017, Sudevan2018, Sudevan2020a, Sudevan2020b, Yadav2021, Sudevan2022}. {\it A major bottleneck to similarly disentangle CMB B mode and foreground components  is the contamination by the strong diffused synchrotron and thermal dust foreground components apart from the thermal noise originating from the circuitry of the measuring detectors. Therefore, it is important to reduce the effects of foreground and detector noise using techniques of machine learning and estimate cleaned PGW signal. We  show that an innovative machine learning technique can remove these non-cosmological contaminations and accurately  estimate pristine CMB B mode map using detailed simulations of $20$  frequency maps  of upcoming  ECHO satellite mission.} Various other applications of machine learning technique in CMB context are also in place~\cite{DipoleANN2023,ClANN2023,Chanda2021,Adams2023,Wang2022, Yan2023,Heinrich2024,Caldeira2019, Petroff2020}.


\begin{figure*}
\includegraphics[scale=0.62]{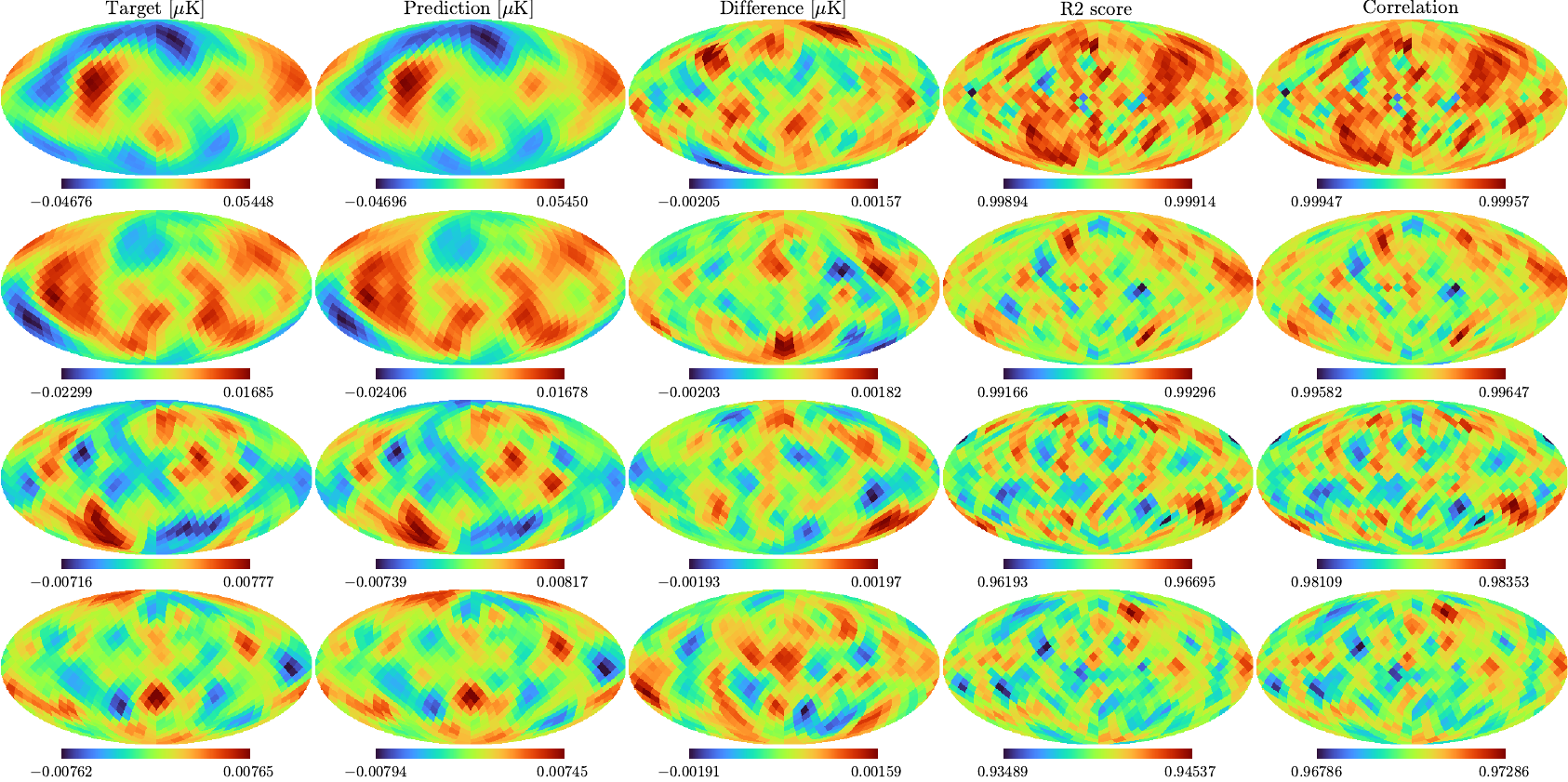}
\caption{\label{MapFigure}Figure showing accurate reconstruction results for CMB B mode maps. From top to bottom each row correspond to the case of $r_{\textrm{test}} = 0.03, 0.003, 0.0003$ and  $0$ respectively. Unit for the first three columns is in $\mu K$ thermodynamic.}
\end{figure*}

{\bf Methodology:} Each training realization comprises of all $20$ frequency maps of ECHO (at HEALPix~\cite{Gorski2005} pixel resolution $N_{\textrm{side}} = 8$) at the input of the PrimeNet. Focusing on a generic network model capable of predicting the pure CMB B mode signal for a sufficiently wide and physically motivated range of variations of $r$ we generate $2000$ random CMB B mode maps for each $r$ taken from a set of total $16$ different  values  equally spaced in $\log$ intervals between $r=0.0001$ and $0.055$. Other cosmological parameters are consistent with~\cite{Planck2020} with lensing amplitude $A =1$~\cite{Camb2000}. The foreground models for training, validation  and testing realizations are created by PySM~\cite{Thorne2017} which consists of all  diffused polarized foregrounds -- anomalous
microwave~\cite{Draine1998,Draine1999,QUIJ2016}, synchrotron and thermal dust emissions. For robust CMB prediction we diversify the samples with a suite of foreground models. The synchrotron models consisted of pixel varying spectral index ($\textrm{S}_1$) and with a frequency break model alongside the spectral index variation ($\textrm{S}_3$) of PySM. For thermal dust we use pixel varying  dust temperature and spectral index model ($\textrm{D}_1$)~\cite{Planck2016} and another version of the two component thermal dust model ($\textrm{D}_4$)~\cite{FDS1999,Meisner2015}. Thus each input realization contains foregrounds identical to  any one of the  $4$ possible sets $\textrm{F}_1 = \bigl(\textrm{A}_2, \textrm{S}_1, \textrm{D}_1\bigr)$, $\textrm{F}_2=\bigl(\textrm{A}_2, \textrm{S}_1, \textrm{D}_4\bigr)$, $\textrm{F}_3=\bigl(\textrm{A}_2, \textrm{S}_3, \textrm{D}_1\bigr)$ and $\textrm{F}_4=\bigl(\textrm{A}_2, \textrm{S}_3, \textrm{D}_4\bigr)$, where $\textrm{A}_2$ represents anomalous microwave component. Each of these foreground model sets is infused with $500$ different CMB realizations for each training $r$ value. For validation realizations we use a different set of $250$ CMB maps for each of the above foreground sets and for each  $r$ taken from the set $r=\bigl(0.02, 0.002, 0.0002\bigr)$. We use a pixel uncorrelated white noise model consistent with ECHO polarized detectors and applicable to $N_{\textrm{side}} = 8$. The CMB, foreground and noise maps at all frequencies of each training and validation realizations are smoothed by a  polarized Gaussian window function of FWHM = $20^\circ$. This causes all angular scales  with $\ell$  larger than $\ell_{\textrm{max}} = 180^\circ/20^\circ = 9$  to become  beam limited. We chose $\ell_{\textrm{max}}=9$ for producing angular spectra. Training, validation and testing sets respectively contain $32000$, $3000$ and $10000$ (for each $r$) independent realizations.

\begin{figure*}
\includegraphics[scale=0.54]{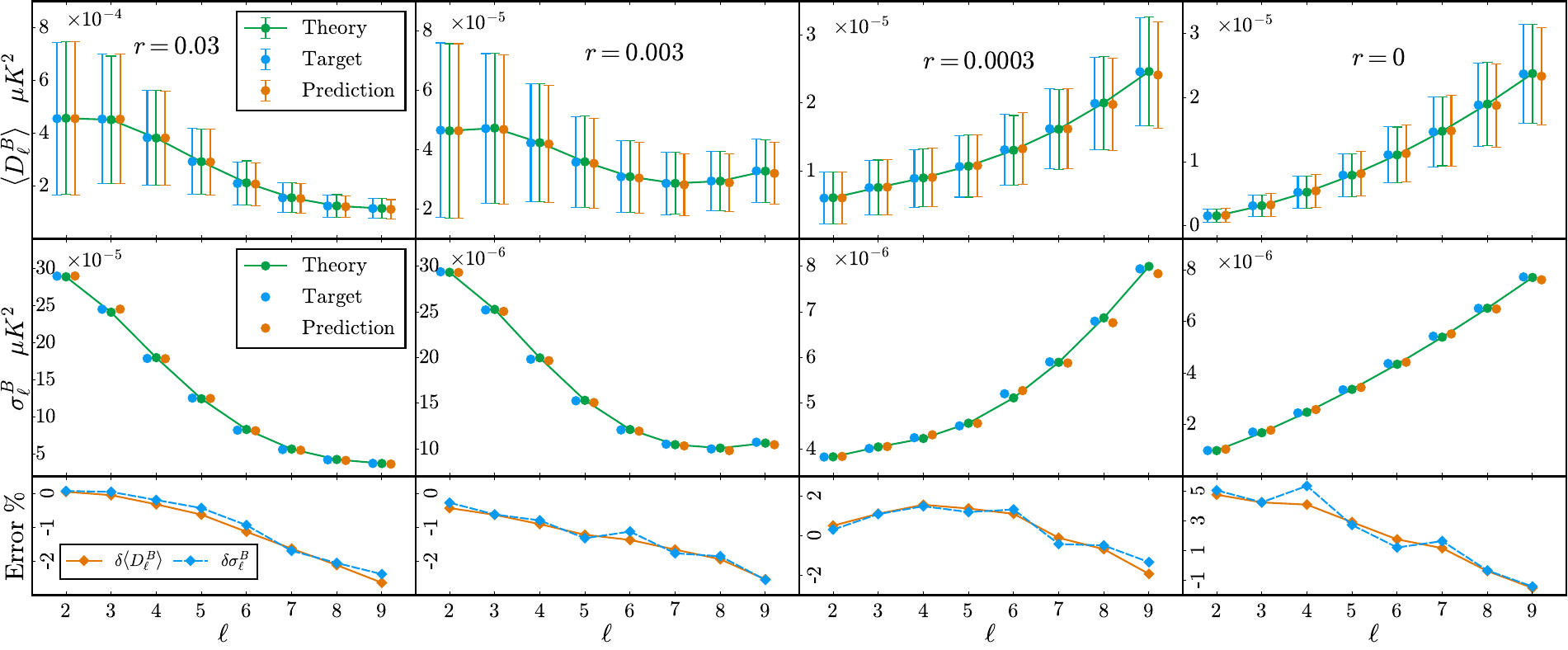}
\caption{\label{ClFigure} Comparison of  angular spectra (top row), their standard deviations (middle) and fractional bias in prediction results  of first two rows (bottom). Value of  $r$ is same for a given column. }
\end{figure*}

The PrimeNet is an autoencoder containing a total of $7$ layers starting from the input through the bottleneck to the final output layers. Subsequent to the input layer the encoder contains three layers with  $192$, $48$ and $12$ nodes corresponding to $N_{\textrm{side}} = 4, 2$ and $1$ respectively. After the bottleneck layer of least pixel resolution the decoder consists of  two decoding layers of $N_{\textrm{side}}=2$  and $4$ respectively. The output layer produces a single map at the original resolution and contains $768$ nodes. Each layer contains $220$ filters. Both the kernel size and stride for all filters are set to  $4$ since filters are designed to reduce sizes of the maps by the same factor  as the pixel downgrading operation of HEALPix. We employ the sigmoid-weighted linear unit~\cite{Silu2017} activation function to learn the non-linearity in the hidden layers of the network. However, we apply linear activation in both the bottleneck and output layers. We use mean absolute error loss function~\cite{Qi2020} for prediction. Before feeding the ECHO realizations to the network we standardize them by first subtracting the mean and  thereafter dividing by the standard deviation~\cite{kotsiantis2018}. This approach facilitates the generalization of network training as well as expedites the deep learning process.

For the backward propagation~\cite{backprop1989,backprop1992}, we utilize the adaptive moment estimation~\cite{adam2014} optimizer with learning rate $0.0001$ to update the weights and biases based upon the loss value calculated in the forward propagation. We  utilize the mini-batch algorithm~\cite{minibatch2016, minibatch2020} with   batch size $64$ and a total of $2000$ epochs.

The performance of the training of the autoencoder can
vary at some level due to the random initialization of realization of the weights in the network. This variation reflects as the epistemic uncertainty in the prediction of the autoencoder. We use model averaging ensemble~\cite{LAI2022} method to reduce this epistemic uncertainty in the prediction.  Final B mode map is obtained by averaging predictions of $60$  models each starting from independent points in the multidimensional weight and bias spaces.



{\bf Results:} For testing case we chose realizations corresponding to $r_{\textrm{test}} = 0.03, 0.003, 0.0003$ and $0$.  These $r$ values were never used in training. Moreover the $r_{\textrm{test}}=0$  denotes the lowest possible value and well below  the minimum $r$ ($0.0001$) used in the training. For each $r_{\textrm{test}}$  we use $60$ ensemble averaged network predictions to evaluate performance for $10000$ test cases.  Again, network inputs for tests contain completely different CMB and noise realizations from those used in training and validation cases.

From top to  bottom each row of Fig.~\ref{MapFigure} summarizes the results for cleaned B map reconstruction for some randomly chosen test realizations with decreasing values of $r_{\textrm{test}}$. First and second column confirm the accurate reconstruction over the largest scales of the maps. The difference maps (third column) contain pixel values significantly less than the target and predictions for each $r_{\textrm{test}}$. An $\textrm{R}_2$ score statistic defined by  one minus the ratio of mean squared residuals between the target and prediction and mean target variance computed at each map pixel over $10000$ test cases is used to quantify the prediction accuracy of cleaned maps. The statistic scores $\gtrsim 99.8\%, 99.1\%, 96.1\%$ for $r_{\textrm{test}} = 0.03, 0.003$ and $0.0003$ respectively, indicating accurate reconstruction by PrimeNet. {\it We emphasize the results of last row in which case the tensor mode was set to zero deliberately. Although, $r=0$ value is well below the minimum $r = 0.0001$ used in the training  we see that PrimeNet predicts the lensed B mode signal accurately, preserving features over large scales (target and predicted maps) with $\textrm{R}_2$ score between $93.4\%$ to $94.5\%$.} Variation of $\textrm{R}_2$ score between $99.7\%$  to $99.9\%$ was obtained in test results with $r_{\textrm{test}} = 0.1$. (This choice of $r$ is well above the maximum $r= 0.055$  used in the training.) {\it This indicates that PrimeNet effectively learned to remove the foreground and noise features from the input maps. This is an excellent feature to achieve since the predictions in this case are robust with respect to major unknowns on the real sky - variations of actual value of $r$ and  complex and realistic foreground models.} Large $\textrm{R}_2$ score values represent reconstruction error between the target and predictions are minimal.  The last column of Fig.~\ref{MapFigure} shows the correlation maps between the target and prediction B mode maps for various values of $r_{\textrm{test}}$. The high correlation values for all cases demonstrate that the predictions accurately follow the target  CMB B  maps. We emphasize on the
striking similarities between the patterns of the last two columns. This strengthens further that the predictions are free from any residual systematic (e.g, due to residual foregrounds).

\begin{figure*}
\includegraphics[scale=0.54]{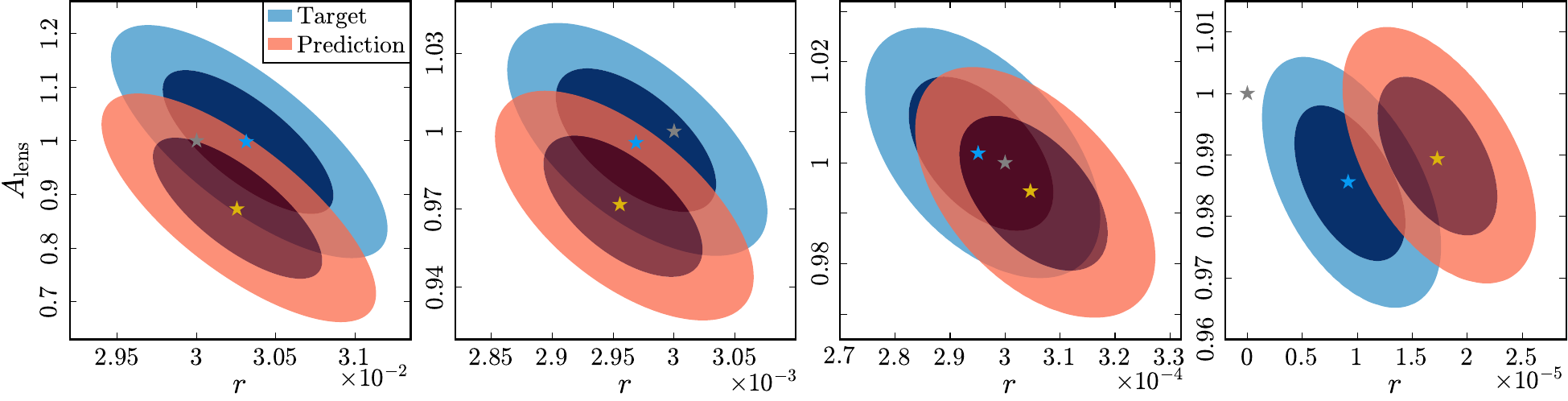}
\caption{\label{likelihood} $1\sigma$ and $2\sigma$ confidence regions for $A$ and $r$ from the mean likelihood function of all $400$ test predictions for $r_{\textrm{test}} = 0.03, 0.003, 0.0003, 0$ from left to right. The stars denote best fit positions. Grey star represents the theoretical parameter values.}
\end{figure*}

The direct predictions of PrimeNet are the CMB maps. How well the angular power spectra of the predicted maps agree the corresponding spectra of the target maps? Accurate production of these spectra is essential for the method to be useful in the estimation of $r$. We see that the spectra of the predicted maps agree excellently with those of the targets for angular scales above the beam resolution of the input maps (i.e, $\ell \lesssim 9$). This underscores high accuracy already achieved in map predictions on the relevant scales and overall consistency of the results. In the following we use the word `predicted (target) spectra' to merely imply the angular spectra estimated from predicted (target) maps while remembering that the actual predictions of PrimeNet are only the cleaned maps.

We compute the mean spectra separately from the predicted and target maps from $10000$ test cases for $r_{\textrm{test}} = 0.03, 0.003, 0.0003, 0.0$ respectively. In the top panel of Fig.~\ref{ClFigure} we show these mean spectra along with the theoretical CMB B mode spectra $D^B_{\ell} = \ell\left(\ell+1\right)C^B_{\ell}/(2\pi)$. The target, predicted and theoretical spectra at each multipole coincide very closely with each other over the vertical scales, however, for visual clarity the spectral locations of these values at each $\ell$ are shifted slightly along the horizontal axis. This demonstrates unbiased reconstruction of angular spectra over the relevant angular scales. The error bars for  predicted spectra agree well with the target (or theoretical cosmic variance). This indicates maximally possible accurate reconstruction limited only by cosmic variance. In the second panel we compare the standard deviations of spectra from predicted, target samples along with cosmic variance limit for all the $r_{\textrm{test}}$ values. All these error estimates agree well. {\it The agreement between the angular spectra, their errors between the target and predictions appear even more remarkable noting that they are not direct predictions. This is a direct consequence of accurate learning by PrimeNet to remove non cosmological signals.}  The third row zooms in the fractional difference of the mean predicted spectra with respect to the mean target spectra. Also shown in this panel are the fractional differences of sample standard deviations of predicted compared to the target. For $r_{\textrm{test}} = 0.03, 0.003, 0.0003$  cases this fractional difference lie within $\left(-2.63\%, 0.04\%\right)$, $\left(-2.53\%, 0.43\%\right)$ and $\left(-1.93\%, 1.56\%\right)$ respectively. For $r_{\textrm{test}} = 0$  the fractional bias lies between $\left(-1.52\%,4.76\%\right)$. These fractional bias values are much less than the fractional cosmic variance induced errors, which are as large as $63.2\%$ at $\ell =2$ and $32.4\%$ at $\ell = 9$. This clearly shows that the predicted spectra are highly unbiased and accurate at the same level. 
Spectral reconstructions are accurate even if  $r_\textrm{test}$ values were not used in training and $r_{\textrm{test}} = 0$ is well below the minimum training $r$. {\it This is a reconfirmation that the PrimeNet learned to remove non-CMB features efficiently and produces robust predictions.}

How well can we measure the primordial $r$ from the predicted B mode maps? The maximum likelihood values of $r$ and $A$ is obtained by maximizing the likelihood function
\begin{equation}
P(r,A) \sim \exp\biggl(-\frac{1}{2}\sum_{\ell=2}^{9}(2\ell+1)\biggl[\frac{\hat C^B_{\ell}}{C^B_{\ell}}+\ln\biggl(\frac{C^B_{\ell}}{\hat C^B_{\ell}}\biggr)-1\biggr]\biggr)\nonumber\,
\end{equation}
where $\hat C^B_{\ell}$ represents prediction for a test case and $C^B_{\ell} =C^B_{\ell}(r,A)$ denotes the theoretical spectrum. To evaluate any possible bias and quantify the accuracy of estimation of $r$ we estimate $P(r,A)$ for $400$ randomly chosen predictions for each $r_{\textrm{test}}$. The product of these $400$ (2-d) likelihood functions (after normalizing peak values to unity) represents the mean likelihood function. We show them in  Fig.~\ref{likelihood}.
The predicted best fit values for $r$ are respectively $0.03025 \pm 0.00024, 0.00296 \pm 3.42 \times 10^{-5}, 0.0003 \pm 7.68\times 10^{-6}, 1.73 \times 10^{-5} \pm 3.06 \times 10^{-6}$ compared with corresponding target values $0.03031 \pm 0.00024, 0.00297 \pm 3.43\times 10^{-5}, 0.000295 \pm 7.42\times 10^{-6}, 9.17 \times 10^{-6}\pm 2.85 \times 10^{-6}$. Ratio of the error estimates for prediction and target sets are close to unity except for $r=0$ in which case we get $\sim 7\%$ larger error for predictions. The prediction and target $r$ obtained from the respective mean likelihood functions differ by merely $0.25, 0.39, 1.25, 2.5$ times the predicted $\sigma$. {\it Both parameters and error limits  agree excellently with those obtained from the mean target. This shows the residual bias in estimation of $r$ are negligible and results are cosmic variance dominated.}

\begin{figure}
\includegraphics[scale=1]{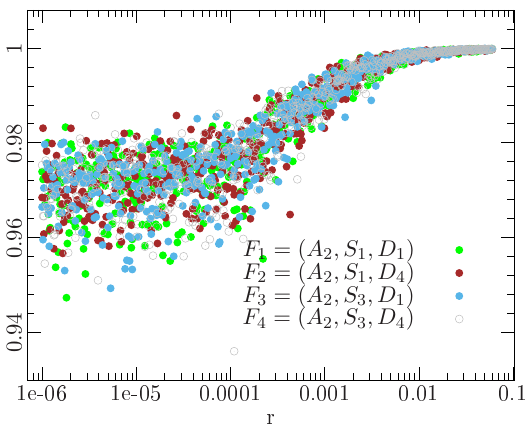}
\caption{\label{corr} Figure showing strong correlation between predicted and target CMB B mode maps for $r \in (10^{-6}, 0.055)$ and for all foreground sets.}
\end{figure}

In addition to  above we test performance by computing correlations between target and predictions, where  input simulated ECHO  maps to PrimeNet contain B mode realizations for $500$ discrete $r$ values chosen to be equally spaced in $\log(r)$ intervals between $r_{\textrm{min}} = 10^{-6}$ and $r_{\textrm{max}}= 0.055 $ for all foreground sets $\textrm{F}_1$ to $\textrm{F}_4$. For each prediction correlation between the target and predicted maps are obtained which are shown in Fig.~\ref{corr}. The predictions possess a strong correlation ($0.936$ to $0.998$) with the targets in all cases. {\it Although the correlation values decreases somewhat when $r\lesssim 0.003$ and tend to spread nominally for $10^{-6}\lesssim r \lesssim 0.0001$ we note that the estimated $r$ values for these cases are still cosmic variance dominated, since in the test case with $r=0$ we demonstrated that the prediction errors  are determined  only by cosmic variance. Therefore, the correlation values of Fig.~\ref{corr}  demonstrate efficient predictions of PrimeNet virtually at any $r$ within interval $\left[10^{-6}, 0.055\right]$ and  even for $r$ values well below the minimum $r = 0.0001$ used in the training.}


{\bf Summary:} For the first time in literature we show that it is possible to extract weak unknown  $r$ value for the PGW from CMB B mode signal over large angular scales of the sky ($ \gtrsim 20^\circ$) based upon training a convolutional neural network (PrimeNet) using all the $20$ frequency maps  of upcoming satellite based CMB polarization mission ECHO.  The low resolution maps were chosen since one is primarily interested in primordial B mode signal containing unique signature from gravitational waves generated from inflation. {\it Extensive tests on PrimeNet show that B mode CMB spectra and tensor to scalar ratio, $r$ estimated from the predicted maps are limited only by the cosmic variance,  the  theoretically minimum uncertainty limit achievable since we have only one Universe to do all the cosmological experiments.}

The input frequency maps for training were created using  the most complex and diverse polarized foreground models produced by PySM which incorporate latest state of the art knowledge about the polarized synchrotron, thermal dust and anomalous microwave components enriched  by the WMAP and Planck observations. Specifically, we include both $\textrm{S}_1$ and $\textrm{S}_3$ models for synchrotron and $\textrm{D}_1$ and $\textrm{D}_4$ models for thermal dust and $\textrm{A}_2$ model for anomalous microwave from PySM. The instrumental noise is assumed to be white with a level consistent with ECHO channels. {\it  Accurate reconstruction in presence of complex and diverse  foreground models in the training configurations ensures robust predictions with model variations.  Detailed and complex foreground modeling enables accurate description of realistic observations. Our network can also be deployed in analysis of other future generation full sky CMB polarization missions.}

The complex foreground models were varied uniformly between sets $\textrm{F}_1$ to $\textrm{F}_4$ in all the training and the testing phases. The detector noise varied  with realizations of  simulated ECHO experiments.  {\it PrimeNet  predicts the B mode  maps, angular spectra over large angular scales with $\ell \lesssim 9$ and hence $r$ with errors given only by the cosmic variance, even in presence of noise. This is a major advantage since one is able to rid of noise (and foreground) effects significantly.} Neural networks for denoising images have been employed by other groups~\cite{diffusion2015, diffusion2020, diffusion2021}.

The training samples contained realizations of CMB B mode with $r$ varying between $0.0001$ to $0.055$ in $16$ equal step in $\log(r)$ interval and a lensing signal with amplitude parameter $A = 1$ in CAMB.  During testing we find that PrimeNet can accurately predict B mode maps even for those $r$ values which were never included in the training. This strongly suggests that learning  of non CMB signals to be removed is achieved efficiently.  {\it Detailed testing with $r_{\textrm{test}} = 0.03, 0.003, 0.0003$ and $0.0$ shows that the maximal likelihood values of predicted $r$ and their uncertainties closely agree with the pure target realizations. Hence our results are effectively only cosmic variance limited.} The predicted maps for $500$ different $r$ values in the interval $10^{-6}$ to $0.055$ show strong correlation with the target maps implying efficient reconstruction  even for  $r \lesssim 0.0001$ which were not used in training.
{\it Thus successful prediction across wide range of foreground models and unseen and lower $r$ cases than used in training provides encouraging results and is a new method to efficient and robust prediction of $r$. This provides a new pathway leading to an open window to the primordial universe.}

\begin{acknowledgments}
SKY acknowledges support by SERB, Government of India, through the National Post Doctoral Fellowship grant (PDF/2022/002449/PMS). SKY acknowledge National Supercomputing Mission (NSM) for providing computing resources of ‘PARAM Porul’ at NIT Trichy, which is implemented by C-DAC and supported by the Ministry of Electronics and Information Technology (MeitY) and Department of Science and Technology (DST), Government of India.
\end{acknowledgments}



\end{document}